\documentclass[twocolumn]{aastex631}

\usepackage{multirow}
\usepackage{CJK}

\shorttitle{Bursting massive protostar M17~MIR}
\shortauthors{Chen, Zhiwei et al.}

\begin{document}
\begin{CJK*}{UTF8}{gbsn}
   \title{M17~MIR: A Massive Protostar with Multiple Accretion Outbursts \footnote{Based on observations collected at the European Southern Observatory, Chile, Programs 073.C-0170(A+B), and 077.C-0174(A).}}

   \correspondingauthor{Zhiwei Chen}
   \email{zwchen@pmo.ac.cn}

   \author[0000-0003-0849-0692]{Zhiwei Chen (陈志维)}
   \affil{Purple Mountain Observatory, Chinese Academy of Sciences, 10 YuanHua Road, 210023 Nanjing, China}
   \author{Wei Sun (孙玮)}
   \affil{Purple Mountain Observatory, Chinese Academy of Sciences, 10 YuanHua Road, 210023 Nanjing, China}
   \author{Rolf Chini}
   \affil{Astronomisches Institut, Ruhr--Universit{\"a}t Bochum, Universit{\"a}tsstrasse 150, 44801 Bochum, Germany}
   \affil{Instituto de Astronom{\'i}a, Universidad Cat{\'o}lica del Norte, Avenida Angamos 0610, Casilla 1280 Antofagasta, Chile }
   \author{Martin Haas}
   \affil{Astronomisches Institut, Ruhr--Universit{\"a}t Bochum, Universit{\"a}tsstrasse 150, 44801 Bochum, Germany}
   \author{Zhibo Jiang (江治波)}
   \affil{Purple Mountain Observatory, Chinese Academy of Sciences, 10 YuanHua Road, 210023 Nanjing, China}
   \author{Xuepeng Chen (陈学鹏)}
   \affil{Purple Mountain Observatory, Chinese Academy of Sciences, 10 YuanHua Road, 210023 Nanjing, China}

  \begin{abstract}

 We report the discovery of a massive protostar M17~MIR embedded in a hot molecular core in M17. The multiwavelength data obtained during 1993--2019 show significant mid-IR (MIR) variations, which can be split into three stages: the decreasing phase during 1993.03--mid-2004, the quiescent phase from mid-2004 to mid-2010, and the rebrightening phase from mid-2010 until now. The variation of the 22\,GHz H$_2$O maser emission, together with the MIR variation, indicates an enhanced disk accretion rate onto M17~MIR during the decreasing and rebrightening phases. Radiative transfer modeling of the spectral energy distributions of M17~MIR in the 2005 epoch (quiescent) and 2017 epoch (accretion outburst) constrains the basic stellar parameters of M17~MIR, which is an intermediate-mass protostar ($M_\ast\sim5.4\,M_\sun$) with $\dot M_\mathrm{acc}\sim1.1\times10^{-5}\,M_\sun\,\mathrm{yr^{-1}}$ in the 2005 epoch and $\dot M_\mathrm{acc} \sim1.7\times10^{-3}\,M_\sun\,\mathrm{yr^{-1}}$ in the 2017 epoch. The enhanced $\dot M_\mathrm{acc}$ during outburst induces the luminosity outburst $\Delta L\approx7600\,L_\sun$. In the accretion outburst, a larger stellar radius is required to produce $\dot M_\mathrm{acc}$ consistent with the value estimated from the kinematics of H$_2$O masers. M17~MIR shows two accretion outbursts ($\Delta t\sim 9-20$ yr) with outburst magnitudes of about 2 mag, separated by a 6\,yr quiescent phase. The accretion ourbusrt occupies 83\% of the time over 26 yr. The accretion rate in outburst is variable with amplitude much lower than the contrast between quiescent and ourbusrt phases. The extreme youth of M17~MIR suggests that minor accretion bursts are frequent in the earliest stages of massive star formation.

 \end{abstract}

   \keywords{Accretion, accretion disks -- Stars: massive -- Stars: formation -- Stars: protostars -- Stars: individual objects (M17~MIR)}

%

\section{Introduction} \label{Sect:Int}
High-mass stars ($M_\ast \gtrsim 8 \,M_\sun$) have strong impacts on their environment through their various feedback affects. However, there is still no consensus on the basic formation mechanism of high-mass stars. The massive star formation models of disk accretion and accumulated observational evidence appear to favor the idea that high-mass stars form as lower-mass progenitors that continuously accrete material from their gas-rich circumstellar disks at high accretion rates ($\gtrsim10^{-4}\,M_\sun\,\mathrm{yr}^{-1}$) \citep[e.g.,][]{2010ApJ...721..478H,2016A&ARv..24....6B,2016A&A...585A..65H,2018ARA&A..56...41M}. Observational evidence for episodic accretion extends across a wide range of protostellar mass, from low values \citep[e.g.,][for a review]{2014prpl.conf..387A} to a high mass $\sim10-20\,M_\odot$ \citep[e.g.,][]{2015MNRAS.446.4088T,2017NatPh..13..276C,2017ApJ...837L..29H}. Although the total periods of multiple bursting accretions constitute only a few percent of the entire accretion phases of massive young stellar objects (MYSOs), bursting accretions can contribute a substantial fraction of the zero-age main-sequence (ZAMS) mass of MYSOs \citep{2019MNRAS.482.5459M}. Moreover, bursting accretions influence the evolutionary tracks of MYSOs; during each bursting accretion, MYSOs experience rapid excursions toward more luminous, but colder, regions of the Hertzsprung-Russel diagram, which are usually occupied by evolved massive stars \citep{2019MNRAS.484.2482M}. This short bloating of MYSOs can occur multiple times up to the end of the pre-main-sequence evolution, different from the spectroscopically confirmed bloating of MYSOs during the contraction phase toward the main sequence \citep{2015A&A...578A..82C,2017A&A...604A..78R}. The discovery of burst-induced photometric variability for MYSOs is impossible at optical wavelengths and difficult in the IR, due to severe extinction and the large distances of MYSOs.

At a distance of $\approx2.0$\,kpc \citep{2011ApJ...733...25X,2016MNRAS.460.1839C,2019ApJ...874...94W}, M17 is among the best laboratories in the Galaxy for investigating the formations of high-mass stars. Several efforts focused on the high-mass stars in M17 have attempted to measure the physical properties of MYSOs, e.g., \citet{1997ApJ...489..698H}, \citet{2001A&A...377..273N}, \citet{2006A&A...457L..29H}, and \citet{2017A&A...604A..78R}. Meanwhile detailed studies of several peculiar MYSOs provide better constraints on their evolutionary stages and circumstellar environments, e.g., \citet{2000A&A...357L..33C,2004Natur.429..155C,2006ApJ...645L..61C}, \citet{2002AJ....124.1636K}, \citet{2007ApJ...656L..81N}, \citet{2015A&A...578A..82C}, and \citet{2020ApJ...888...98L}. Apart from the classified MYSOs (mostly IR-bright), it is very likely that unknown high-mass protostars are deeply embedded in dense cores within M17. There are nine sites of 22\,GHz H$_2$O masers in M17 \citep{1998ApJ...500..302J,2010MNRAS.406.1487B}; however, only one is likely associated with the known massive protostar IRS5A \citep{2015A&A...578A..82C}; the driving sources of the remaining H$_2$O masers are still unclear.

The expanding shell traced by the H$_2$O maser spots \citep[][hereafter CJO16]{2016MNRAS.460.1839C} motivates the search for the driving source. The coordinates of the putative driving source are determined as $\mathrm{R.A.(J2000)}=18^\mathrm{h}20^\mathrm{m}23\fs017, \mathrm{decl.(J2000)}=-16\degr 11\arcmin 47\farcs98$ with positional errors of the order of 1\,mas (CJO16), which coincides with a faint point source (hereafter M17~MIR) with Spitzer $4.5$ and $5.8\,\mu$m magnitudes of 9.97 and 7.82 mag, respectively (see more descriptions of GLIMPSE in Sect.~\ref{Sect:spitzer}). On the other hand, none of the compact radio sources reported by \citet{2012ApJ...755..152R} matches M17~MIR, which eliminates the association with a hyper- or ultracompact \ion{H}{2} region, and thus indicates a very early stage for M17~MIR.

This paper reports the analyses of the multiwavelenth data available for M17~MIR. The multiwavelength data are described in Sect.~\ref{Sect:Obs}. The analyses of these data and the results are presented in Sect.~\ref{Sect:Res}. The variability of M17~MIR is classified in Sect.~\ref{Sect:var}. The derived properties of M17~MIR are discussed in Sect.~\ref{Sect:Dis}, and the conclusions are summarized in Sect.~\ref{Sect:Con}.


\begin{figure*}
\centering
\includegraphics[width=1.0\textwidth]{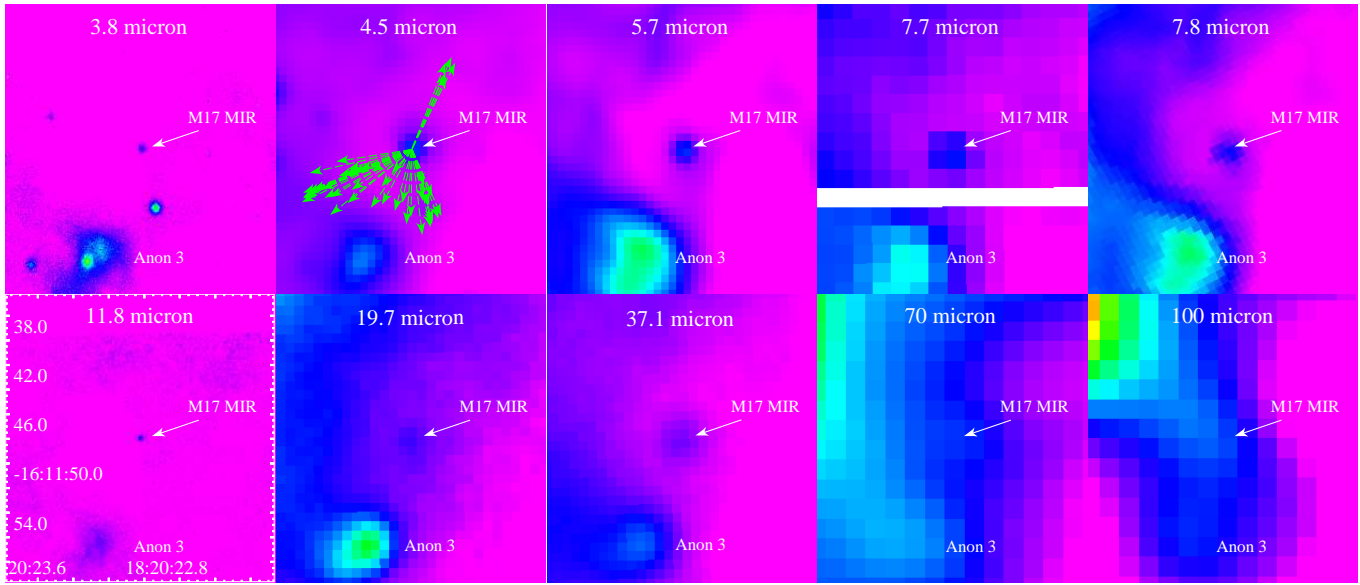}
\caption{Multiwavelength images centered on M17~MIR. Proper motions of H$_2$O maser spots (CJO16) are overlaid on the image at $4.5\,\mu$m.}
\label{Fig:img}
\end{figure*}

\section{Observations}\label{Sect:Obs}
\subsection{IR Imaging Data from the Very Large Telescope}
The wide-field $JHK_s$- and $L'$-band images were obtained with ISAAC at the Very Large Telescope (VLT) in 2002 September and 2004 May to September, respectively. The pixel scale is $0\farcs148$ and $0\farcs07$, and the FWHM is $0\farcs6$ and $0\farcs5$ for the $JHK_s$- and $L'$-band images, respectively \citep[see more details of observations in][]{2008ApJ...686..310H}. These $JHK_sL'$-band images and photometric data were published by \citet{2008ApJ...686..310H}.

The MIR images were obtained toward the M17\,SW photodissociation region (PDR) \citep[e.g.,][]{1999A&A...352..277C,2012A&A...542L..13P} with the VLT imager and spectrometer for mid-infrared \citep[VISIR;][]{2004Msngr.117...12L} at the VLT in 2006 April through the SiC filter ($\lambda=11.8\,\mu$m, $\Delta \lambda=2.1\,\mu$m). The data were taken with the \textsf{VISIR\_img\_obs\_AutoChopNod} template. The photometric standard star was HD\,178345 during this observation. The data were reduced and calibrated using the EsoReflex workflows \citep{2013A&A...559A..96F} and the VISIR pipeline recipes 4.3.7\footnote{http://www.eso.org/sci/software/pipelines/}. The image is of good quality with FWHM $\approx 0\farcs4$, reaching the diffraction limit.




\subsection{The Infrared Space Observatory (ISO) Data\footnote{Based on observations with ISO, an ESA project with instruments funded by ESA Member States (especially the PI countries: France, Germany, the Netherlands and the United Kingdom) and with the participation of ISAS and NASA.}}
The archival imaging data of ISO \citep{1996A&A...315L..27K} were obtained for the M17 SW PDR from the ISO Data Archive\footnote{https://www.cosmos.esa.int/web/iso/access-the-archive}. The ISO observations for the M17 SW PDR were carried out in 1997 March and published in \citet{1999A&A...352..277C}; they aim at the interstellar medium inside the PDR. The calibrated ISO/ISOCAM images in the three broad filters (LW1, LW4, and LW6) with central wavelengths of 4.5, 6.0, and 7.7\,$\mu$m are used in this paper.

\subsection{The Spitzer Data}\label{Sect:spitzer}
The Galactic Legacy Infrared Mid-Plane Survey Extraordinaire \citep[GLIMPSE;][]{2003PASP..115..953B,2009PASP..121..213C} is a legacy science program of the Spitzer Space Telescope that surveyed the inner Galactic plane in the 3.6, 4.5, 5.8 and $8.0\,\mu$m bands with the IR camera IRAC. The angular resolution in the four bands (I1,I2,I3,and I4) is about $2\farcs5$. The GLIMPSE mosaicked images of M17 were obtained from the NASA/IPAC Infrared Science Archive\footnote{http://irsa.ipac.caltech.edu/frontpage/}; the median observation date was 2005 September. The GLIMPSE point sources were retrieved from the GLIMPSE I Spring 07 Archive\footnote{See more descriptions in http://irsa.ipac.caltech.edu/Missions/spitzer.html}.

M17 was revisited by Spizter in the warm phase in 2017 July. During these observations, only the two bluest bands ($3.6$ and $4.5\,\mu$m) were available, and the flux-calibrated images (program ID: 13117, observer: Robert Benjamin) were obtained from the Spitzer Heritage Archive\footnote{Hosted at https://irsa.ipac.caltech.edu/Missions/spitzer.html}.

\subsection{The Herschel Data}

In order to cover the far-IR regime, we incorporated the publicly available imaging observations from the Herschel Space Observatory (Herschel) and its PACS instrument at 70, 100, and 160\,$\mu$m \citep{2010A&A...518L...2P}. The Herschel data used in this work are Level 3.0 products (eight observations with IDs from 1342192767 to 1342192774) provided by the Herschel Science Archive\footnote{http://archives.esac.esa.int/hsa/whsa/}. The pixel scales are $1\farcs6$, $1\farcs6$, and $3\farcs2$ at 70, 100, and 160\,$\mu$m, and the measured resolutions are $8\farcs0$, $8\farcs0$, and $12\farcs0$, respectively.

\subsection{The WISE/NEOWISE data}
The Wide-field Infrared Survey Explorer \citep[WISE,][]{2010AJ....140.1868W} is a 40\,cm telescope in a low earth orbit that surveyed the entire sky in 2010 using four IR bands at 3.4, 4.6, 12, and 22\,$\mu$m. The angular resolution in the four bands (W1, W2, W3, and W4) are $6\farcs1$, $6\farcs4$, $6\farcs5$, and $12\farcs0$, respectively. Only the WISE W1 and W2 data of M17 are considered in this work, because the M17 \ion{H}{2} region is saturated in the W3 and W4 bands.  With the primary aim of studying near-Earth Objects, NEOWISE observations resumed in 2013 December and continue to date, after the telescope's cryogen tanks were depleted. Due to the orbit of the telescope, WISE/NEOWISE observations visit M17 twice a year, in March and September. From the NEOWISE 2020 data release containing W1 and W2 observations from 2013 December until 2019 December, the single-exposure data of M17 collected over 6 yr (12 epochs) are analyzed in this work, in addition to the WISE all-sky survey image of M17 (two epochs).

The single-exposure images of the WISE/NEOWISE mission toward M17 were retrieved from the WISE data hosted on IRSA \footnote{https://irsa.ipac.caltech.edu/Missions/wise.html}, with the service of WISE/NEOWISE Coadder. The single exposures were coadded to obtain higher-quality W1/W2 images with enhanced resolution of $0\farcs6875/\mathrm{pix}$. Finally WISE/NEOWISE images from 14 epochs of M17 were used in this work. Aperture photometry with a fixed radius of 8.5 pixel ($=5\farcs8$) is applied to all WISE/NEOWISE images with a field of view of $3\arcmin\times3\arcmin$. Because the WISE/NEOWISE W2 band is very similar to the Spitzer I2 band, the photometric results from multi epochs are checked against the isolated star G015.0198-00.6768 located $33\arcsec$ south of the image center. G015.0198-00.6768 is not cataloged in the ALLWISE catalog \citep{2014yCat.2328....0C} or in the unWISE catalog \citep{2019ApJS..240...30S}, but has a cataloged Spitzer I2 magnitude of 8.55 \citep{2009yCat.2293....0S}.

\subsection{SOFIA Data}
The SOFIA\footnote{SOFIA is jointly operated by the Universities Space Research Association, Inc. (USRA), under NASA contract NAS2-97001, and the Deutsches SOFIA Institut (DSI) under DLR contract 50 OK 0901 to the University of Stuttgart.} \citep{2012SPIE.8444E..10Y} observations toward the M17 SW PDR were carried out with the FORCAST instrument \citep{2013PASP..125.1393H} at $19\,\mu$m ($\lambda_\mathrm{eff}=19.7\,\mu$m; $\Delta \lambda=5.5\,\mu$m) and $37\,\mu$m ($\lambda_\mathrm{eff}=37.1\,\mu$m; $\Delta \lambda=5.5\,\mu$m) on 2017 August 2 (PI: James M. De Buizer). The calibrated Level 3 images at 19 and $37\,\mu$m were downloaded from the SOFIA Data Cycle System \footnote{https://dcs.arc.nasa.gov/}. The resolutions at 19 and $37\,\mu$m are $3\farcs3$ and $3\farcs5$, respectively \citep{2017ApJ...843...33D}. The SOFIA/FORCAST images fill the gap between the VLT/VISIR and Herschel images in wavelength coverage.

\subsection{JCMT SCUBA2 Submillimeter data }
The James Clerk Maxwell Telescope (JCMT) SCUBA2 submillimeter observations were taken at $450\,\mu$m and $850\,\mu$m on 2016 April 10 for M17, as a part of the JCMT large program ``SCOPE: SCUBA-2 Continuum Observations of Pre-protostellar Evolution'' (JCMT program code: M16AL003). The pipeline-reduced and calibrated maps of M17 at $450\,\mu$m and $850\,\mu$m were retrieved from Canadian Astronomy Data Centre \footnote{http://www.cadc-ccda.hia-iha.nrc-cnrc.gc.ca/en/jcmt/}.

\begin{figure*}[!t]
  \centering
  \includegraphics[width=0.8\textwidth]{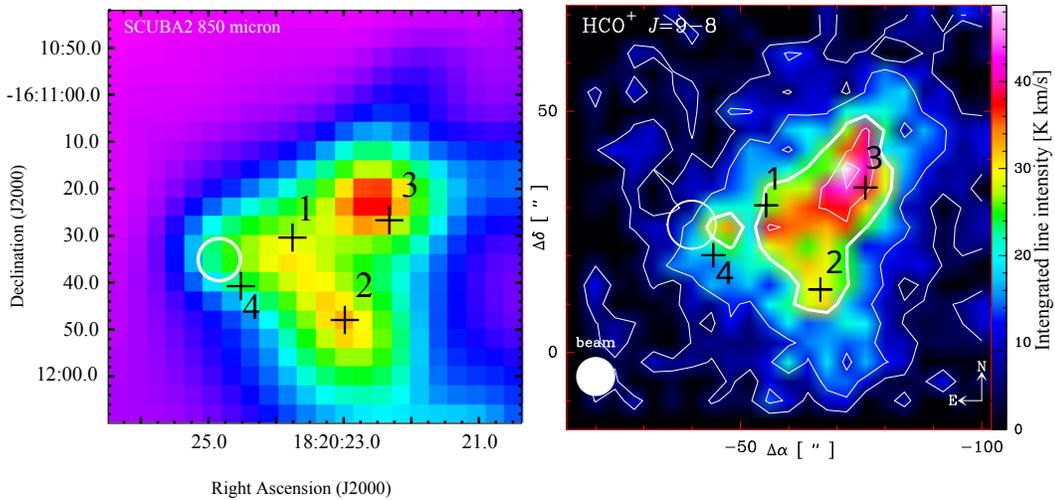}
  \caption{JCMT-SCUBA2 850 $\mu$m map of M17 SW in left, and the HCO+ $J=9\rightarrow8$ integrated intensity map of M17 SW in right \citep{2015A&A...583A.107P} . The 22\,GHz water maser sources detected by \citet{1998ApJ...500..302J} are denoted by crosses with numbers as same as \citet{1998ApJ...500..302J}. The white circle in both maps denote the position of M17 UC1. }
  \label{Fig:HMC}

\end{figure*}

\section{Results}\label{Sect:Res}

\subsection{A very early  high-mass protostellar object driving a protostellar outflow}\label{Sect:HM}

The earliest detection of M17~MIR was at $3.9\,\mu$m $L$-band by the Canada--France--Hawaii Telescope (CFHT) observations in 1992 December and 1993 May with a spatial resolution of $1\arcsec$ \citep{1994A&A...291..239G}. This object, named as source 2 by the authors, and six other extremely red objects are suggested to be young stars in very early evolutionary stages \citep[see Table 1 in][]{1994A&A...291..239G}. Figure~\ref{Fig:img} presents the multiwavelength images centered on M17~MIR. A summary of the IR flux densities obtained by the various facilities except for the earliest CFHT observation is presented in Table~\ref{tbl:IR}.

The source sizes in the various IR images are in proportional to FWHMs of the corresponding images. In the $11.8\micron$ image of $0\farcs4$ resolution ($=800\,\mathrm{au}$), M17~MIR is still a single source. M17~MIR appears as a compact point source at $3-37\,\micron$. At Herschel $70\micron$ and $160\micron$, the far-IR emission is dominated by the contamination from M17~IRS5 and M17~UC1 to the northeast. M17~MIR is not resolved at its IR location of M17~MIR, likely due to the strong ambient far-IR emission. For M17~MIR, we estimated an upper limit $<77\,\mathrm{Jy}$ at $70\,\micron$. 

The faint IR brightness of M17~MIR suggests that it is a low-luminosity object or the extinction to M17~MIR is quite high. From a quick examination of the IR images of M17~MIR, we think high extinction is the more likely reason. M17~MIR's brightness shows a steep rise in the $3-8\,\micron$ range (see also Figure~\ref{Fig:R17}). However, M17~MIR is not visible in the ISO broad filter image at $9.6\,\micron$ \citep[see Figure~2a--f;][]{1999A&A...352..277C}. We know that the silicate absorption feature at $9.7\,\micron$ can cause high and broad absorption to a highly obscured object at around this wavelength. The high extinction to M17~MIR can explain the nondetection at $9.6\,\micron$.

\setcounter{table}{0}
\begin{table}
\centering
\caption{IR flux densities}
\label{tbl:IR}
\begin{tabular}{l c c  c c  }
\hline\hline
Telescope &  $\lambda_\mathrm{eff}$  & Epoch &  Flux Densities & FWHM  \\
                    &   ($\mu$m)  &   & (mJy) &  (arcsec)\\
\hline
VLT & 3.78 & 2004.09 & $1.91\pm0.17$ & 0.5 \\
Spitzer     & 4.50 & 2005.07 & $18.4\pm2.2$ & 2.0 \\
Spitzer     & 5.71 & 2005.07  & $85.0\pm8.0$ & 2.0 \\
ISO        & 7.72 & 1997.03  & $(2.2\pm0.3)\times10^2$ & 3.7  \\
Spitzer    & 7.85 & 2005.07 & $(1.4\pm0.3)\times10^2$ &  2.0 \\
VLT     & 11.80 & 2006.04 & $(1.4\pm0.07)\times10^2$ & 0.4 \\
SOFIA       & 19.7 & 2017.08 & $(5.1\pm1.2)\times10^2$ & 3.3\\
SOFIA   & 37.1 & 2017.08 & $(7.3\pm1.2)\times10^3$ & 3.5 \\
Herschel & 70 &  2010.03  & $<7.7\times10^4$ & 8.0 \\
\hline
\end{tabular}
\tablecomments{Central wavelengths and effective bandwidths are adopted from the SVO Filter Profile Service \citep{2020sea..confE.182R}.}
\end{table}

 The IR flux densities of M17~MIR are at least two orders of magnitude lower than those of the IR-bright high-mass protostar IRS5A \citep{2015A&A...578A..82C} within the same M17 SW PDR. On the other hand, the outwards motion of the CJO16 H$_2$O masers very likely traces the shocked gas accompanying the protostellar outflow driven by M17~MIR. In the earliest stage of high-mass protostars, the mass loss might evolve from a highly uncollimated expanding shell to a collimated outflow \citep[e.g.,][]{2015Sci...348..114C}. The dynamical age of the expanding shell driven by M17~MIR is 12.5 yr (CJO16), younger than the dynamical age ($\sim25$ yr) of W75N(B)-VLA\,2 \citep{2015Sci...348..114C}. The faint IR brightness of M17~MIR and the young dynamical age of the CJO16 H$_2$O masers suggest that M17~MIR is a promising candidate for a high-mass protostar in the earliest stage. The analysis of the spectral energy distribution (SED) constructed from Table~\ref{tbl:IR} will help to better constrain the fundamental properties of M17~MIR (Sect.~\ref{Sect:SED}).

The warm area close to the central object emits radiation mostly at MIR wavelengths, while the surrounding envelope of much larger size is expected to radiate at submillimeter to millimeter wavelengths due to a much lower effective temperature. Figure~\ref{Fig:HMC} presents maps of the $850\mu$m dust continuum emission and the HCO+ $J=9\rightarrow8$ line integrated intensity of M17 SW PDR. M17~MIR, represented by the H$_2$O maser source 2 detected by VLA observations \citep[][hereafter JDG98]{1998ApJ...500..302J}, is located exactly in the secondary peak of the 850\,$\mu$m dust continuum emission. This dust continuum emission peak, along with the other two peaks, make up the northern condensation in M17 SW PDR that was first noticed by \citet{1990ApJ...356..513S}. The three main components in the northern condensation in M17 SW PDR were studied in detail through submillimeter and millimeter observations. Their physical properties were estimated by \citet{1993MNRAS.264.1025H} (their Table 4). M17~MIR is located within the FIR3 source in \citet{1993MNRAS.264.1025H}. On the other hand, the map of integrated line intensity (range: $0-40\,\mathrm{km\,s^{-1}}$, peak velocity $\sim+20\,\mathrm{km\,s^{-1}}$) of the HCO+ $J=9\rightarrow8$ line shows clumpy structures similar to the dust continuum emission. A dense core whose size is comparable to the beam size of the observations ($\sim8\arcsec$, or 0.08 pc) is coinciding with M17~MIR. Given the critical density $n_\mathrm{crit}=9\times10^7\,\mathrm{cm}^{-3}$  and an excitation temperature $E_\mathrm{up}=192.58\,\mathrm{K}$ of the $802.458\,\mathrm{GHz}$ HCO+ $J=9\rightarrow8$ line \citep{2015A&A...583A.107P}, the dense core associated with M17~MIR is likely a candidate for a hot molecular core \citep[HMC;][]{2005IAUS..227...59C}. HMCs commonly corresponds to a hydrogen column density $N_\mathrm{H}\gtrsim10^{23}$ or even higher \citep[e.g.][]{2017A&A...604A..60B}. Although a high-mass protostar embedded in an HMC is undetectable at short wavelengths due to high extinction, the detection of the central protostar becomes possible at MIR wavelengths. For instance, the IR counterpart F4 of the HMC G9.62+0.19-F appears at $3.8\,\mu$m in the $L'$ band and at longer wavelengths \citep{2005A&A...429..903L}.

The multiwavelength data of M17~MIR and the submillimeter to millimeter data of its surrounding medium indicate that M17~MIR is a very early high-mass protostar embedded in an HMC and is also driving a protostellar outflow.

\setcounter{table}{1}
\begin{table}
  \tabletypesize{\normalsize}
\centering
\caption{$L$-band magnitudes of embedded YSOs in M17~SW during the 1993.03 and 2004.09 epochs.}
\label{tbl:l}
\begin{tabular}{c c c c c }
\hline\hline
Name & R.A.  & Decl.   &  1993.03  & 2004.09 \\
     &  (J2000) & (J2000) & (mag)   & (mag)     \\
\hline
2 = M17~MIR & 18:20:23.02 & -16:11:47.9 & $10.2$ & $12.78$  \\
3 & 18:20:23.91 & -16:11:43.4 & $10.8$ & $10.38$ \\
5    & 18:20:24.10 & -16:11:40.1  & $11.2$ & $10.99$ \\
7    & 18:20:22.34 & -16:11:28.7  & $8.8$  & $8.76$  \\
\hline
\end{tabular}
\tablecomments{See Figure~\ref{Fig:spz} for the positions of these sources on the Spitzer I1 and I2 images.}
\end{table}

\section{Variability}\label{Sect:var}

\subsection{Infrared variability}\label{Sect:IR_var}

For M17~MIR, the photometric magnitude in the 1993 CFHT observation was $10.2\pm0.2$ \citep[source 2 in Table~2;][]{1994A&A...291..239G}, while the magnitude in the 2004 VLT observation was $12.78\pm0.14$ \citep{2008ApJ...686..310H}. The filter profiles in the two observations are almost identical. Table~\ref{tbl:l} lists the $L$-band magnitudes of the embedded YSOs in the two observations. Interestingly, M17~MIR is about 10 times fainter in the 2004.09 epoch than in the 1993.03 epoch, whereas the other sources are fairly stable.

The VLT $L'$-band observations in 2004.09 were one year earlier than the Spitzer observations in 2005.09. By interpolating the Spitzer flux densities at I2 and I3 to that at $L'$ under the assumption of a blackbody radiation, the flux density of M17~MIR in the $L'$ band is about 2.2\,mJy in the 2005.09 epoch. This predicted flux density in 2005.09 is close to the $1.91\pm0.17$ mJy measured in 2004.09. We therefore speculate that M17~MIR's luminosity is stable during 2004.09--2005.09.

\begin{figure}[!t]
\centering
\includegraphics[width=0.45\textwidth]{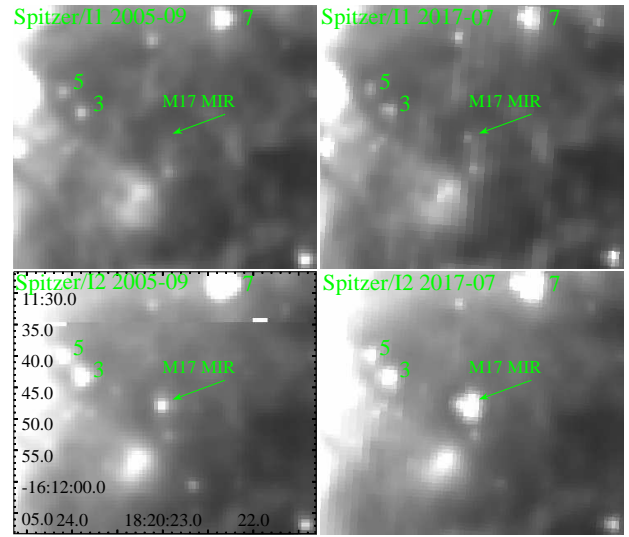}
\caption{Top panel: Spitzer/I1 images in 2005.09 (left) and 2017.07 (right). Bottom panel: Spitzer/I2 images in 2005.09 (left) and 2017.07 (right). The embedded YSOs 3, 5, and 7 are the same as in Table~\ref{tbl:l}.  }
\label{Fig:spz}
\end{figure}

M17~MIR shows variabilities in the Spitzer I1 and I2 images from 2005.09 and 2017.07 (see Figure~\ref{Fig:spz}). Comparing the flux densities at I1 in the two epochs for an aperture radius of $1\farcs6$, we find an increase by a factor of 2.9. Similarly, the flux densities at I2 show an increase by a factor of 3.3 in an aperture radius of $3\farcs2$. In contrast, the other sources are stable in the two bands (see Table~\ref{tbl:spz}).

\begin{figure}[!t]
\centering
\includegraphics[width=0.45\textwidth]{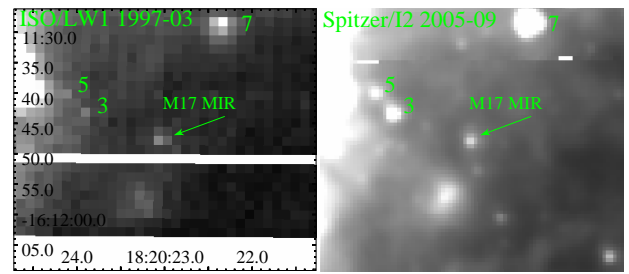}
\caption{ISO/LW1 image in 1997.03 (left) and Spitzer/I2 image in 2005.09 (right). The embedded YSOs 3, 5, and 7 are identical to those in Figure~\ref{Fig:spz}. }
\label{Fig:iso}
\end{figure}

The MIR variabilities of M17~MIR found for different epochs motivate us to compare its flux densities from the ISO observations in 1997.03 and the Spitzer observations in 2005.09 and 2017.07. Table~\ref{tbl:iso} shows the flux densities of M17~MIR and other nonvariable sources in the ISO/LW1 and Spitzer/I2 bands, which have almost identical filter profiles centered at $4.5\,\mu$m. Figure~\ref{Fig:iso} compares the ISO LW1 and Spitzer I2 images obtained in 1997.03 and 2005.09, respectively. Direct comparison from Figure~\ref{Fig:iso} clearly indicates that M17~MIR is significantly brighter than sources 3 and 5 in the ISO LW1 image in the 1997.03 epoch, while it becomes fainter than sources 3 and 5 in the Spitzer I2 image in the 2005.09 epoch. Given that sources 3 and 5 are unchanged (see context above), M17~MIR is varying between 1997.03 and 2005.09. The aperture photometry of M17~MIR in the $4.5\,\mu$m band shows variability from 1997 to 2017.

\setcounter{table}{2}
\begin{table}
  \tabletypesize{\normalsize}
\centering
\caption{Flux densities of M17~MIR Measured from the Spitzer Observations in 2005 and 2017}
\label{tbl:spz}
\begin{tabular}{c c c c c c }
\hline\hline
Name & Spitzer &   2005  & 2017  & Aperature \\
     &   Band           &     (mJy)   & (mJy)   &  (arcsec) \\
\hline
\multirow{2}{1.3cm}{M17~MIR}   & I1         &  $1.0\pm0.4$  &   $2.9\pm0.5$  & 1.6\\
     & I2         &  $19.0\pm2.5$ &   $52.9\pm4.7$ & 3.6 \\
  \multirow{2}{1.3cm}{3}          & I1          &   $9.7\pm0.9$            & $9.6\pm0.8$ &  2.0              \\
     & I2         &   $33.8\pm3.7$             &   $33.7\pm3.9$    & 2.5     \\
  \multirow{2}{1.3cm}{5}          & I1          &  $10.7\pm0.7$    & $8.3\pm0.6$     & 2.0           \\
     & I2         &  $23.8\pm1.5$               &  $20.9\pm1.3$    & 2.0       \\
  \multirow{2}{1.3cm}{7}          & I1          &  $37.8\pm5.0$   & $31.3\pm3.7$  & 2.7       \\
                                  & I2          &  $201.2\pm17.2$  & $193.0\pm6.9$   & 4.0        \\
\hline
\end{tabular}
\tablecomments{Aperture photometry on the sources in this table uses fixed in positions, aperture size, and circular annulus size in the 2005 and 2017 epochs, for direct comparison between the two epochs.}
\end{table}

\setcounter{table}{3}
\begin{table}
  \tabletypesize{\normalsize}
\centering
\caption{Flux densities of M17~MIR at $4.5\,\mu$m measured with ISO/ISOCAM in 1997 and Spitzer in 2005 and 2017.}
\label{tbl:iso}
\begin{tabular}{c c c c c c c }
\hline\hline
  Name &  1997  & 2005    & 2017  & Aper. \\
       & (mJy)  & (mJy)   & (mJy) &  ($\arcsec$) \\
\hline
M17~MIR   &   $32.1\pm11.1$    & $19.0\pm2.5$   &  $52.9\pm4.7$   &  3.6 \\
 7          &  $202.7\pm31.9$    & $201.2\pm17.2$ &     $193.0\pm6.9$           &  4.0 \\
\hline
\end{tabular}
\tablecomments{The aperture photometry of M17~MIR in the ISO/LW1 image might be affected by the row of bad pixels lying below M\,17~MIR, thus its flux density might be underestimated. }
\end{table}

\begin{figure*}[!hbt]
  \centering
  \includegraphics[width=0.99\textwidth]{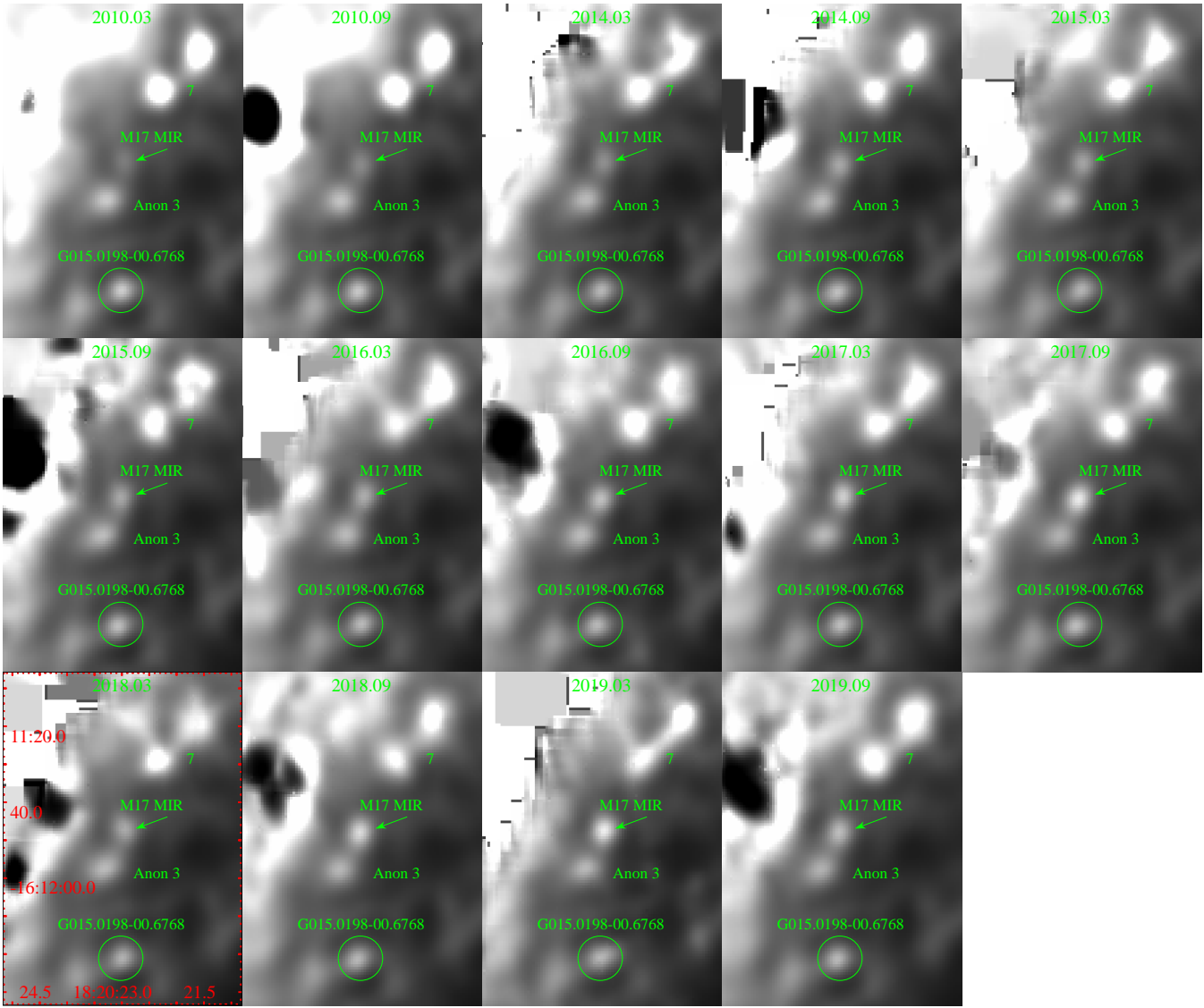}
  \caption{WISE/NEWOWISE W2 images centered on M17~MIR in 14 epochs.}
  \label{Fig:wise_img}

\end{figure*}

For M17, WISE/NEOWISE images in 14 epochs, including 2010.03 and 2010.09, and March and September in 2014 to 2019, were analyzed in this work. M17~MIR is only detected in the W2 band. The WISE W1 flux densities, extrapolated from its flux densities in the Spitzer I1 and I2 bands, are about 0.6 and 1.8\,mJy during the 2005.09 and 2017.07 epochs, respectively. The extreme faintness at W1 is consistent with the nondetection of M17~MIR in the corresponding W1 images.

Figure~\ref{Fig:wise_img} presents the WISE/NEOWISE W2 images centered on M17~MIR in 14 epochs. The pixel resolution of $0\farcs6875$ clearly separates M17~MIR from the nearby IR source Anon 3 to the southwest. The photometric results of M17~MIR, source 7, and G015.0198-00.6768 are tabulated in Table~\ref{tbl:wise} for 14 epochs. In 2010.03, the W2 flux density of M17~MIR was $17.7\pm4.4$ mJy, which is very close to the value $18.4\pm2.0$ in the Spitzer I2 band in 2005.09. Given the fact that the WISE/NEOWISE W2 band is very similar to the Spitzer I2 band, M17~MIR was roughly stable in brightness from 2004 to 2010. However, six months later M17~MIR was twice as bright in 2010.09, with a flux density of $33.7\pm4.4$ mJy. It is very likely that the rebrightening of M17~MIR started between 2010.03 and 2010.09. In the observations of NEOWISE, M17~MIR generally shows a steady rise in brightness, but with a jump and fall motion epoch by epoch in its light curve (see Figure~\ref{Fig:lvar}). Between 2016 and 2017, M17~MIR underwent a steep rise in brightness with an amplitude of 1.7. In 2018 M17~MIR's brightness dropped back to the level of 2016. and was followed by a rise with a mean amplitude of 1.5 between 2018 and 2019.

\subsection{Combined light curve in the $L'$ and W2 bands}\label{Sect:mir_lc}

\begin{figure}[!t]
\centering
\includegraphics[width=0.48\textwidth]{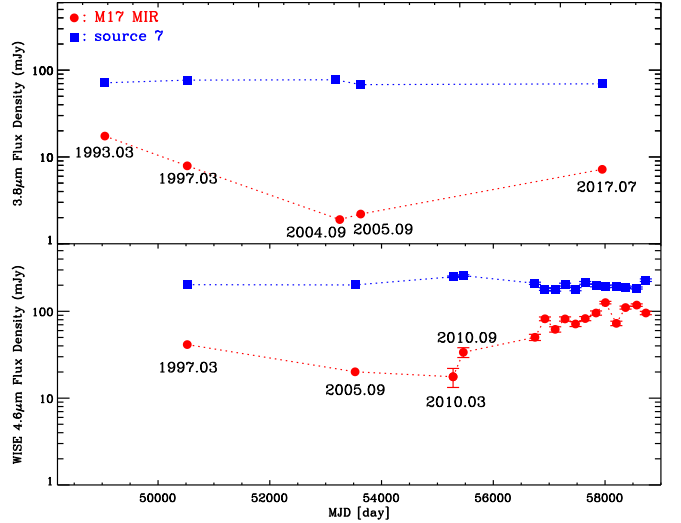}
\caption{Top: light curve of M17~MIR at $3.8\,\mu$m in the $L'$ band; Bottom: light curve at $4.6\,\mu$m in the W2 band. The time span is over 26 yr from 1993 to 2019. }
\label{Fig:lvar}
\end{figure}

The MIR flux measurements of M17~MIR span several wavelengths and observation epochs. The $3.8\,\mu$m $L'$ filter is closest to the $L$ filter of the 1993.03 epoch observation \citep{1994A&A...291..239G}, while the observations of other epochs provided flux densities in at least two filters. For a direct comparison from 1993 to 2017, the flux density in the $3.8\,\mu$m $L'$ filter is estimated for 1993.03, 1997.03, 2005.09, and 2017.07. The multiwavelength flux densities of M17~MIR in 1997.03, 2005.09, and 2017.07 epochs yield a reasonable estimate for the continuum in the range $3-8\,\mu$m, assuming blackbody emission. The flux density of M17~MIR in the $3.8\,\mu$m $L'$ filter is the continuum emission inside the transmission curve of the $L'$ filter. Note that the blackbody temperatures for the three different epochs are different. For the $L$-band observations in 1993, the three blackbody temperatures are used for estimating the $L'$ flux density of M17~MIR. We find that the largest difference among the three values is less than 15\%, thus the mean value is used as the $L'$ filter flux density. For 1993.03, 1997.03, 2005.09, and 2017.07 we obtain $L'$ flux densities; for 2004.09 there is direct $L'$-band photometry for M17~MIR. Figure~\ref{Fig:lvar} presents the combined light curve of M17~MIR in the $L'$ and W2 bands. The time span of the multiepoch data is 26.5 yr.

M17~MIR shows decreasing, quiescent and rebrightening phases in its MIR light curve, as shown in Figure~\ref{Fig:lvar}. Due to the lack of $3-4\,\mu$m observations before 1993 and during 1997--2004, the duration of the decreasing phase is poorly constrained. From the available data, the quiescent phase between the decreasing and rebrightening phase might last about 6 yr, from mid-2004 to mid-2010. Because M17~MIR seems to continue the trend of rebrightening, the timescale of the rebrightening phase is $\gtrsim9$ yr. The continuing NEOWISE observations will provide vital information on the rebrightening timescale.

\setcounter{table}{4}
\begin{table}
  \tabletypesize{\normalsize}
\centering
\caption{WISE/NEOWISE flux densities of M17~MIR and other IR sources}
\label{tbl:wise}
\begin{tabular}{c c c c c  }
\hline\hline
  Epoch & MJD &  M17~MIR  & source 7    & G015.0198 \tablenotemark{a}  \\
        &     &    W2 (mJy)       &  W2 (mJy)    &   W2 (mJy)                  \\
\hline
2010.03	&	55,280.5 	&	17.6 	(4.4)	&	250.0 	(5.1)	&	70.7 	(2.4)	\\
2010.09	&	55,463.5 	&	33.7 	(4.4)	&	257.8 	(5.7)	&	72.4 	(2.3)	\\
2014.03	&	56,745.5 	&	50.2 	(4.1)	&	210.3 	(6.4)	&	71.9 	(2.3)	\\
2014.09	&	56,924.6 	&	82.0 	(4.1)	&	180.5 	(6.4)	&	72.0 	(2.4)	\\
2015.03	&	57,107.5 	&	62.0 	(4.4)	&	177.3 	(6.4)	&	69.9 	(2.4)	\\
2015.09	&	57,284.5 	&	81.7 	(4.6)	&	203.3 	(7.7)	&	69.0 	(3.2)	\\
2016.03	&	57,471.5 	&	71.5 	(4.7)	&	180.2 	(7.5)	&	68.9 	(2.8)	\\
2016.09	&	57,645.5 	&	82.5 	(4.5)	&	212.8 	(7.7)	&	66.3 	(3.3)	\\
2017.03	&	57,838.5 	&	95.6 	(5.3)	&	199.4 	(7.7)	&	69.2 	(2.9)	\\
2017.09	&	58,006.5 	&	126.0 	(4.2)	&	193.4 	(6.3)	&	72.2 	(2.4)	\\
2018.03	&	58,202.6 	&	72.7 	(4.3)	&	195.0 	(5.8)	&	67.3 	(2.4)	\\
2018.09	&	58,366.5 	&	110.5 	(4.4)	&	189.2 	(6.3)	&	70.0 	(3.5)	\\
2019.03	&	58,566.5 	&	118.0 	(3.9)	&	185.7 	(5.1)	&	74.9 	(1.6)	\\
2019.09	&	58,730.5 	&	95.7 	(4.2)	&	228.5 	(7.8)	&	68.3 	(2.4)	\\

\hline
\end{tabular}
\tablenotetext{a}{The Spitzer I2 magnitude of this source is 8.55, corresponding to 68.05 mJy, while its WISE/NEOWISE W2 mean flux density is 70.49 mJy. }
\end{table}


\setcounter{table}{5}
\begin{deluxetable*}{ccccccccc}
  \tabletypesize{\footnotesize}
\centering
\tablecaption{Water masers observations}\label{tbl:masers}
\tablewidth{0pt}
\tablehead{
\colhead{Epoch} & \colhead{Obs.} &\colhead{R.A.}      & \colhead{Decl.}       & \colhead{Sep.}      & \colhead{$V$ peak}                 & \colhead{$S$ peak} & \colhead{Beam}  & \colhead{Ref.} \\
  \colhead{}   &      \colhead{}            & \colhead{(J2000)}  & \colhead{(J2000)} & \colhead{(arcsec)}  & \colhead{($\mathrm{km\,s^{-1}}$)} &  \colhead{(Jy)} &   \colhead{}  &   \colhead{} }
\decimalcolnumbers
\startdata
  1984.06 & VLA-C   &  18:20:22.91 &-16:11:48.29 & 1.7 & 23.32 & 1.5 & $4\farcs1\times2\farcs2$  &  FC89 \\
  1984.06 & VLA-C & 18:20:23.12 &-16:11:48.29 & 1.5 & 18.05 & 6.7 & $4\farcs1\times2\farcs2$  &FC89 \\
  1984.06 & VLA-C  &  18:20:23.05 &-16:11:48.20 & 0.5 & 16.73 & 6.9 & $4\farcs1\times2\farcs2$ & FC89 \\
  1984.08 & VLA-D & 18:20:23.07 & -16:11:49.2 & & 17.4 & 18.57 & $5\farcs6\times7\farcs1 $ & MCF88 \\
  1996.09 & VLA-D  &  18:20:22.98 &-16:11:47.97 & 0.6 & 15.8  & 121 & $5\farcs0\times3\farcs4$ &JDG98 \\
  1996.09 & VLA-D  &  18:20:22.98 &-16:11:47.97 & 0.6 & 27.7  & 17  & $5\farcs0\times3\farcs4$ &JDG98 \\
\hline
  2003.10   & ATCA-EW352   &  18:20:23.04 &-16:11:48.4  & 0.6 & 19   & 197 & $8\arcsec\times29\arcsec$  &B10 \\
  2003.10   & ATCA-EW352  &  18:20:23.04 &-16:11:48.4  & 0.6 & 24    & 60  & $8\arcsec\times29\arcsec$  &B10 \\
  2004.07   & ATCA-H168  &  18:20:23.04 &-16:11:48.4  & 0.6 & 20    & 67  & $13\arcsec\times9\arcsec$  &B10 \\
  2004.07   & ATCA-H168  &  18:20:23.04 &-16:11:48.4  & 0.6 & 24    & 17  & $13\arcsec\times9\arcsec$  &B10 \\
  2008.08  & ATCA-6B    &  18:20:23.02 & -16:11:47.8 & 0.2    & 21 & 4.2 & $1.7\times0.5$ & BE11 \\
\enddata
\end{deluxetable*}

\subsection{H$_2$O  Maser Variability}\label{Sect:water}
Table~\ref{tbl:masers} summarizes the 22\,GHz H$_2$O maser observations from 1984 to 2008. The earliest H$_2$O maser detection associated with M17~MIR was achieved at the VLA in C configuration in 1984 June \citep[][hereafter FC89]{1989A&A...213..339F}. The complete catalog of H$_2$O masers found by \citet{1989A&A...213..339F} was compiled by \citet{1999A&AS..137...43F}. Three FC89 H$_2$O masers are associated with M17~MIR, and show two velocity components of $16-18\,\mathrm{km\,s^{-1}}$ and $23.3\,\mathrm{km\,s^{-1}}$. One H$_2$O maser coinciding with M17~MIR was detected with the VLA in D configuration and shows a flux density of 18.57\,Jy \citep[][hereafter MCF88]{1988A&A...194..116M}. One JDG98 H$_2$O maser (source 2 in JDG98) with two velocity components is detected around the same source. The velocities of the FC89 H$_2$O masers are consistent with those of the two components of the JDG98 H$_2$O maser source 2. Both the FC89 and JDG98 H$_2$O masers are likely pumped by the interaction between the outflow from M17~MIR and the surrounding gas. Interestingly, the flux density of the JDG98 H$_2$O maser source 2 in the 1996.09 epoch is ten times higher than that of the FC89 H$_2$O maser in 1984.06.

Several years after the last VLA observations, three water maser observations were taken by ATCA during three epochs. A fast fading of the water maser intensity from 197\,Jy in 2003.10 \citep[][hereafter B10]{2010MNRAS.406.1487B}, through 67\,Jy in 2004.07 (B10), to 4.2\,Jy in 2008.08 \citep[][hereafter BE11]{2011MNRAS.416..178B} was observed. In the earlier two epochs, the spectra of water maser emission show two components-- a stronger blueshifted one at $\sim19-20\mathrm{km\,s^{-1}}$ and a weaker redshifted one at $24\mathrm{km\,s^{-1}}$, which are consistent with the VLA observations in the earlier epochs. In the 2008.08 epoch, two components were observed in the spectrum -- a weaker blueshifted one at $\sim18\,\mathrm{km\,s^{-1}}$ and a stronger redshifted one at $21\,\mathrm{km\,s^{-1}}$.

The latest water maser observations were conducted by very long baseline interferometry (VLBI) with VLBI Exploration of Radio Astrometry (VERA) in 2009.10--2010.12 and consist of six epochs (CJO16). The VERA observations detected 203 water maser spots in these observations. The dominant blueshifted maser spots ($V_{lsr} \lesssim\,22\,\mathrm{km\,s^{-1}}$) are mostly distributed to the south of M17~MIR, while a few redshifted maser spots ($V_{lsr} \gtrsim\,22\,\mathrm{km\,s^{-1}}$) are distributed north to the north of it (see Figure~\ref{Fig:m}). This feature is consistent with the stronger blue-shifted component and the weaker red-shifted component observed by the VLA and ATCA.






\section{Discussion}\label{Sect:Dis}

\subsection{The Connection between the IR Variability and H$_2$O Maser Emission Variability}

The IR variability of M17~MIR overlaps with the surrounding H$_2$O maser variability in time. The connection between them helps to clarify the physical origin of these variabilities. The flux density of the JDG98 H$_2$O maser source 2 in 1996.09 is ten times higher than that of the FC89 H$_2$O maser source in 1984.06. This H$_2$O maser variation from 1984 to 1996 and the decreasing phase in the IR light curve during 1993 and 2004 suggest that the brightness peak likely occurred between 1984 and 1993. Moreover, the fast fading of the H$_2$O maser emission from 2003 to 2008 and the quiescent phase in the IR light curve during 2004 and 2010 together constrain the occurrence of the quiescent phase to between mid-2004 and mid-2010, characterized by both low IR brightness and H$_2$O maser emission.

VERA observations in the earliest two epochs (2009.10 and 2010.01) were made in the quiescent phase, while the later observations in the epochs 2010.04, 2010.09, 2010.11, and 2010.12 took place in the rebrightening phase. The proper motions of the H$_2$O masers spots are traced in more than three epochs (J. O. Chibueze 2021, private communication). The proper motions of the H$_2$O maser spots are measured in the transient phase between quiescent and rebrightening phases from 2009.10 to 2010.12. The 3D motions of the H$_2$O maser spots detected by the VERA observations show an expanding bubble originating from M17~MIR. Together with the rebrightening, we suggest that both phenomena are due to the accretion/outflow activity of M17~MIR.

Similar flaring of the H$_2$O masers associated with the outflow from the high-mass protostar NGC\,6334I-MM1B shows a mean factor of 6.5 before and after the burst of MM1B in the submillimeter range \citep{2018ApJ...866...87B}. Those authors attribute the enhanced accretion rate to the coincident flaring in the submillimeter continuum and the H$_2$O maser emission of NGC\,6334I--MM1B. For M17~MIR, we suggest an accretion outburst to explain both the MIR variations and the correlated H$_2$O maser variations. Accompanied by the expanding motions of H$_2$O maser spots measured during the transient phase from quiescent to rebrightening and the close positive relation between mass loss and accretion rate, the most promising explanation for the outburst in MIR brightness since 2010.03 is an enhanced accretion rate.

\subsection{Outflow rate and disk accretion rate of M17~MIR}\label{Sect:rates}

\begin{figure}[!t]
\centering
\includegraphics[width=0.45\textwidth]{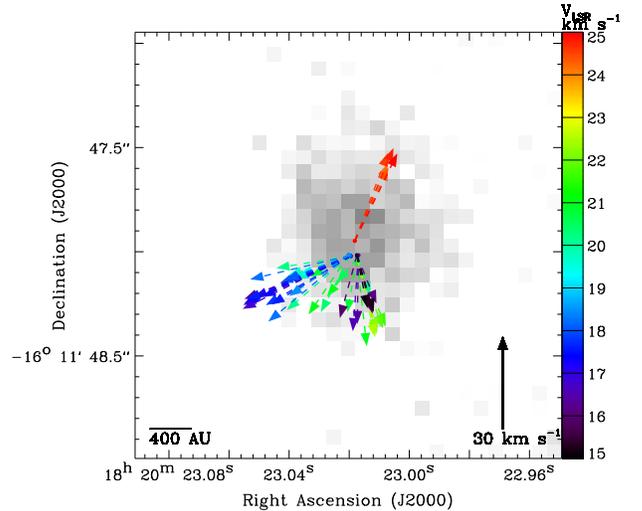}
\caption{The proper motions and $V_\mathrm{LSR}$ of the CJO16 H$_2$O maser. The background grayscale image is the VLT $L'$ image.}
\label{Fig:m}
\end{figure}

The proper motions and radial velocities of the CJO16 H$_2$O masers are overlaid on the VLT $L'$ image, as shown in Figure~\ref{Fig:m}. The 3D motions of the H$_2$O maser spots can be used to estimate the momentum rate of the gas flow, given the assumption that all the momentum of the ejected gas is transferred to the surrounding molecular environment. The momentum rate $\dot P$ is then estimated from the relation \citep{2011A&A...535L...8G}
$$
\dot P = 1.5\times10^{-3}\,V_{10}^2\,R_{100}^2\,(\Omega/4\pi)\,n_8\,M_\odot\,\mathrm{yr^{-1}\,km\,s^{-1}},
$$
where $V_{10}$ is the mean maser velocity in units of $10\,\mathrm{km\,s^{-1}}$, $R_{100}$ is the average distance of H$_2$O masers from the driving source in units of 100\,au, $\Omega$ is the solid angle of the gas flow, and $n_8$ is the gas volume density in units of
$10^8\,\mathrm{cm^{-3}}$. $R_{100}$ is calculated as 0.85 by averaging the distances of the H$_2$O masers from the dynamical center, which is determined at the position of $\mathrm{R.A.(J2000)}=18^\mathrm{h}20^\mathrm{m}23\fs 017, \mathrm{decl.(J2000)}=-16\degr 11\arcmin 47 \farcs 98$ (CJO16). The gas density is about $10^8\,\mathrm{cm^{-3}}$ according to the discussion in Sect.~\ref{Sect:HM}. $\Omega$ is about $4\pi$ for an expanding bubble. The above physical properties from observations yield a momentum rate $\dot P\sim4\times10^{-3}\,M_\odot\, \mathrm{yr^{-1}\,km\,s^{-1}}$ for the protostellar outflow of M17~MIR. This is comparable to outflows of $10^{-3} - 10^0\,M_\odot\, \mathrm{yr^{-1}\,km\,s^{-1}}$ driven by MYSOs  \citep{2016A&A...585A..71M}, and one magnitude higher than the outflow from a low-mass YSO in IRAS\,20231+3440 \citep{2017MNRAS.469.4788O}.

\begin{figure*}
 \centering
 \includegraphics[width=0.9\textwidth]{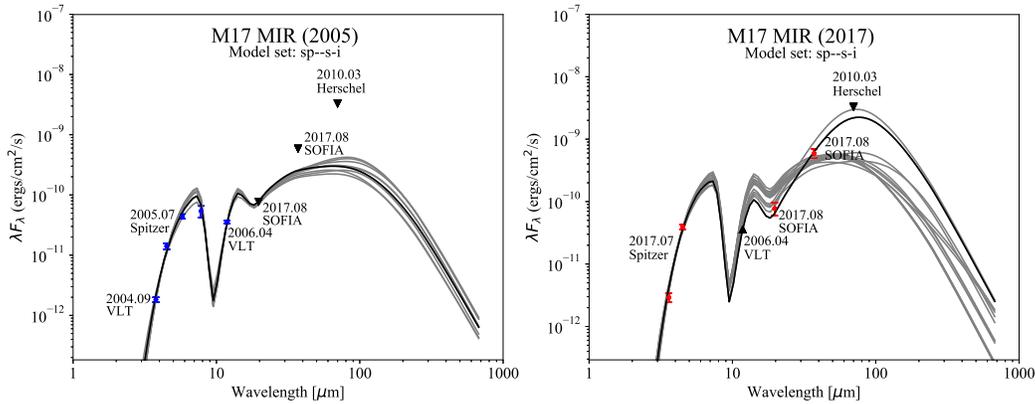}
 \caption{SEDs of the R17 \emph{sp--s-i} models that satisfy $\chi^2 - \chi_\mathrm{best}^2<3 n_\mathrm{data}$ for M\,17~MIR.}
 \label{Fig:R17}
\end{figure*}

The protostellar outflow provides an indirect measure of the disk accretion rate. The outflow mass rate is derived by dividing the outflow momentum rate $\dot P$ by the outflow velocity. Given $\dot P\sim4\times10^{-3}\,M_\odot\, \mathrm{yr^{-1}\,km\,s^{-1}}$ estimated in this work and $V_{10}\approx1.9\,\mathrm{km\,s^{-1}}$ determined by CJO16, the outflow mass rate $\dot M_{out}$ is $\sim2\times10^{-4}\,M_\odot\,\mathrm{yr^{-1}}$. Assuming a ratio of outflow mass rate to disk accretion rate of 30\% \citep{2016A&ARv..24....6B}, the disk accretion rate onto M17~MIR is $\dot M_\mathrm{acc} \sim 7\times10^{-4}\,M_\odot\, \mathrm{yr^{-1}}$.

This disk accretion rate is estimated from the H$_2$O maser observations during 2009 to 2010, when M17~MIR just started to rebrighten. The MIR emission between 2010.09 and 2010.03 varies by a factor of about 2 in the WISE W2 band. \citet{2013ApJ...765..133J} simulate the MIR flux variation induced by an enhanced disk accretion rate and find that the MIR flux variation is proportional to the disk accretion rate. The W2 flux between 2010 and 2019 varies by an average factor of 3. We speculate that the disk accretion rate in 2019 is likely 3 times the rate estimated above, reaching $\sim2\times10^{-3}\,M_\odot\,\mathrm{yr^{-1}}$.



\subsection{Modeling of the SED in Quiescent and Rebrightening phases}\label{Sect:SED}

During the quiescent phase from mid-2004 to mid-2010 (hereafter the 2005 epoch), we collect IR flux densities at six wavelengths to construct the SED of M17~MIR. During the rebrightening phase, the flux densities obtained in 2017.07 (Spitzer I1 and I2 bands) and 2017.08 (SOFIA/FORCAST $19\,\mu$m and $37\,\mu$m bands) are separated by 14 days. The NEOWISE W2 flux density was obtained at the end of 2017 September, nearly two months later than the Spitzer and SOFIA observations. We collect only the Spitzer and SOFIA observations in 2017.07/08 to construct the SED of M17~MIR during the outburst phase (hereafter the 2017 epoch). The flux densities of M17~MIR in the 2005 and 2017 epochs are tabulated in Table~\ref{tbl:sed}. The $3\sigma$ upper limit of the Herschel $70\mu$m image is 77 Jy for M17~MIR, which is considered as the upper limit in both the 2005 and 2017 epochs. The additional input parameters besides the flux densities are distance and foreground extinction $A_V$. The distance to M17~MIR is fixed to 2.0\,kpc. We assume a wide range of 10--200\,mag for the foreground extinction $A_V$.

\setcounter{table}{6}
\begin{table*}
  \tabletypesize{\normalsize}
\centering
\caption{Flux densities of M17~MIR in the 2005 and 2017 epoch}
\label{tbl:sed}
\begin{tabular}{c c c c c c c c c c  }
\hline\hline
Epoch  & Spitzer & VLT & Spitzer & Spitzer & Spitzer &  VLT\tablenotemark{a}  & SOFIA\tablenotemark{b} & SOFIA\tablenotemark{b} & Herschel\tablenotemark{c} \\
   State               & $3.6\mu$m & $3.78\mu$m  & $4.5\mu$m & $5.71\mu$m & $7.8\mu$m & $11.8\mu$m  &$19.7\mu$m & $37.1\mu$m &  $70\mu$m  \\
 &                &(mJy) & (mJy) & (mJy) & (Jy) & (Jy)  & (Jy) & (Jy) &  (Jy) \\
 \hline
  2005 & ... & $2.29\pm0.21$ & $21.3\pm2.9$ & $85.0\pm8.0$ & $0.14\pm0.03$ & $0.139\pm0.007$ & $<0.51\pm0.12$ & $<7.3\pm1.2$ &  $<77$\\
  2017 & $3.8\pm0.7$ & ... & $59.7\pm5.9$ &... & ...  &  $>0.139\pm0.007$  & $0.51\pm0.12$  &  $7.3\pm1.2$ & $<77$ \\
\hline
\end{tabular}
\tablenotetext{a}{The flux density at $11.8\mu$m in the 2005 epoch is considered as the lower limit in the 2017 epoch.}
\tablenotetext{b}{The flux densities at $19.7\mu$m and $37.1\mu$m in the 2017 epoch are considered as the upper limits in the 2005 epoch.}
\tablenotetext{c}{The measurement at $70\mu$m in the 2010.03 epoch is considered as the upper limit.}
\end{table*}

\setcounter{table}{7}
\begin{table*}
  \tabletypesize{\normalsize}
\centering
\caption{$\chi^2$ and $P(D|M)$ scores of R17 model sets}
\label{tbl:fit}
\begin{tabular}{c c c c c c }
\hline\hline
  &   2005 epoch    &  &  & 2017 epoch & \\
  Model & $\chi^2$  & $P(D|M)$ &  Model & $\chi^2$  & $P(D|M)$ \\
  \hline
  sp\--s-i  & 9.034  & 0.0044  &  sp\--s-i & 2.41  & 0.0013  \\
  sp\--smi  & 8.241  & 0.0022  &  sp\--smi & 15.99 & 0.0017  \\
  spu-smi   & 14.54  & 0.0012  &  spu-smi  & 14.46 & 0.0009  \\
  spubsmi   & 9.42   & 0.00065 &  spubsmi  & 11.14 & 0.0005  \\
  \hline
\end{tabular}
\end{table*}

We employed the SED model grid for YSOs that spans a wide range of evolutionary stages, from the youngest deeply embedded protostars to pre-main-sequence stars with debris or no disk, and which covers a wide and uniform region of parameter space \citep[][hereafter R17]{2017A&A...600A..11R}. We fit the SED of M17~MIR using several R17 model sets of different model components, using the Python-based fitting tool \citep[1.3 version;][]{robitaille_thomas_2017_235786}. Based on the prior knowledge of M17~MIR, it is a highly embedded massive protostar with an accretion disk. The model sets we used are sp\--s-i,sp\--smi,spu-smi,spubsmi. All four model sets have a passive disk with an inner radius equal to the dust sublimation radius. sp\--s-i is the simplest model set including only the central star and passive disk. The other three model sets include ambient gas (sp\--smi), envelope+ambient gas (spu-smi), and envelope+ambient+cavity (spubsmi). 

Fitting the SED of M17~MIR to the individual models of the four model sets returns a $\chi^2$ value for each individual model. The criterion $\chi^2 - \chi_\mathrm{best}^2<3 n_\mathrm{data}$ can generally split good and bad fits. We used this criterion to select good fits from the four model sets. Among these good fits, we use the following criteria to pick reasonable models.

Disk mass $\geqslant\,0.01M_\sun$. The disk accretion of M17~MIR estimated from H$_2$O maser motions is $7\times10^{-4} \,M_\sun\,\mathrm{yr}^{-1}$. To sustain such a high accretion rate, we require this criterion. 

Disk inclination angle $\geqslant30\degr$. Because the blue- and redshifted H$_2$O maser spots are distinctly separated in the vicinity of M17~MIR, the disk inclination angle should not be small.

Exclude unphysical models. Some models are unphysical with too small radii but too high temperatures. We compute the Stefan--Boltzmann luminosity ($4\pi R_\ast^2\sigma T_\ast^4$) from the radius and temperature of each model. These unphysical models are below the ZAMS track in the H-R diagram. We keep only the models that are lying on/above the ZAMS track in the H-R diagram.

The above criteria for the four model sets return reasonably good fits for the SEDs of M17~MIR in the 2005 and 2017 epochs. Because of the degeneracy of the four model sets, one has to score these sets not only based on $\chi^2$ values. In order to assess the model sets in comparison with each other in a statistically robust way, \citet{2017A&A...600A..11R} suggested calculating $P(D|M) \varpropto N_\mathrm{good}/N$, where $N_\mathrm{good}$ is the total number of models from a given model set, and $N$ is the total number of models in that set. \citet{2019ApJ...875..135T} used the R17 model sets to find the best YSO models for 12 MYSOs based on the overall assessment of $\chi^2$ and $P(D|M)$ scores. For the good fits of M17~MIR, we show the $\chi_\mathrm{best}^2$ and $P(D|M)$ scores of the four model sets in Table~\ref{tbl:fit}. In the 2005 and 2017 epochs, sp\--s-i is the "best representation" model set according to both the $\chi_{best}^2$ and $P(D|M)$ scores.

Among the sp\--s-i model set, 44 and 13 models remain for the SEDs of M17~MIR in the 2005 and 2017 epochs, respectively. In the 2017 epoch, the SOFIA observations at $19.7\micron$ and $37.1\micron$ are very useful in distinguishing the various models of a given model set (Figure~\ref{Fig:R17}). Among the 13 models in the 2017 epoch (Table~\ref{tbl:R17}), two models correctly reproduce the SED slope between $19.7\micron$ and $37.1\micron$. Indeed, the $\chi^2_\mathrm{best}$ model is different from the second fit only in the disk inclination angle. Aided by the observations at $19.7\micron$ and $37.1\micron$ in the 2017 epoch, the $\chi^2_\mathrm{best}$ model is distinctly better than the other sp\--s-i models. 
    
In the 2005 epoch, the 44 sp\--s-i models show a degeneracy in the long-wavelength SED beyond $12\micron$, due to the lack of observations at long wavelengths. \citet{2011ApJS..194...14P} found similar SED degeneracy at long wavelengths for a deeply embedded YSO with hard X-ray emission (PCYC source 699 from \citet{2011ApJS..194...14P}) in the Carina nebula region, when they fitted the SED of this object to YSO model SEDs from \citet{2006ApJS..167..256R}. In Figure~9 of \citet{2011ApJS..194...14P}, the distributions of the well-fit YSO models show two distinct groups: intermediate-mass YSOs with relatively high circumstellar extinction and low-mass YSOs with lower circumstellar extinction. The group with high circumstellar extinction is consistent with the high absorption of the X-ray source \citep{2011ApJS..194...14P}. For the 44 sp\--s-i models of M17~MIR in the 2005 epoch, two distinct families are found in the plot of infrared luminosity integrated between 12 and $500\micron$ versus disk maximum radius $R_\mathrm{max}^\mathrm{disk}$: high infrared luminosity with $ R_\mathrm{max}^\mathrm{disk}> 1000\,\mathrm{au}$ and lower infrared luminosity with $ R_\mathrm{max}^\mathrm{disk}< 1000\,\mathrm{au}$. The $ R_\mathrm{max}^\mathrm{disk}> 1000\,\mathrm{au}$ family is more consistent with the general understanding of MYSOs, which are luminous at long infrared wavelengths and have large circumstellar disks. Moreover, the $\chi^2_\mathrm{best}$ sp\--s-i model in the 2017 epoch also suggests a large circumstellar disk for M17~MIR. The $ R_\mathrm{max}^\mathrm{disk}> 1000\,\mathrm{au}$ family of the sp\--s-i models is preferred for the SED of M17~MIR in the 2005 epoch. Among the nine sp\--s-i models with $ R_\mathrm{max}^\mathrm{disk}> 1000\,\mathrm{au}$ in Table~\ref{tbl:R17}, two models with $T_{eff}=24,540\,\mathrm{K}$ likely produce significant 22 GHz continuum emission that could be detected by the ATCA observations during 2003--2008 (B10,BE11). However, no 22 GHz continuum source was reported around M17~MIR according to the ATCA observations. We excluded the two very hot sp\--s-i models in the following analysis. On the other hand, stellar luminosity in the 2005 epoch should be much lower than that in the 2017 epoch. This constraint leads to a final collection of only three reasonable sp\--s-i models. For the three sp\--s-i models, we calculated the mean $\langle T_\mathrm{eff}\rangle$, $\langle R_\ast\rangle$,$\langle L_\ast \rangle$, and foreground $\langle A_V\rangle$, weighted by $\frac{1}{\chi^2}$. These values are listed in Table~\ref{tbl:lumin} for the 2005 epoch. 

In the above analysis for the R17 models satisfying $\chi^2 - \chi^2_\mathrm{best}<3n_\mathrm{data}$, no constraint on the foreground extinction $A_V$ was applied. The good SED models in the 2005 and 2017 epochs converge on a narrow range of foreground extinction $A_V = 120-150$ mag, despite a broad $A_V$ range of $10-200$ mag used in the SED fitting procedure. This narrow $A_V$ range still holds for the other ``bad'' SED models. The observations at $7.8\micron$ and $11.8\micron$ in the 2005 epoch that cover the broad absorption feature at $9.7\micron$ yield a strong constrain on the foreground extinction. This is consistent with the deep absorption feature at $9.7\,\mu$m caused by the large absorption of silicate grains, which is commonly observed for deeply embedded high-mass protostars \citep[e.g.,][]{2009ApJ...698..488M,2018ApJ...867L...7P}.


\setcounter{table}{8}
\begin{deluxetable*}{cc cccccccc}
\centering
\tablecaption{The Parameters of R17 sp\--s-i Models that Satisfy $\chi^2 - \chi_\mathrm{best}^2<3 n_\mathrm{data}$ for M\,17~MIR \label{tbl:R17}}
\tablewidth{0pt}
\tablehead{
\colhead{Epoch}   & \colhead{Model Name} & \colhead{$\chi^2$} & \colhead{$\mathrm{A_V}$} & \colhead{$R_\ast$}   & \colhead{$T_\mathrm{eff}$} & \colhead{$M_\mathrm{disk}$} & \colhead{$R_\mathrm{max}^\mathrm{disk}$}    &  \colhead{Inclination} & Comments  \\
          &                     &                    & \colhead{(mag)}         & \colhead{($R_\odot$)} & \colhead{(K)}             & \colhead{($M_\odot$)}       & \colhead{(AU)}                    & \colhead{($\degr$)}}
\colnumbers
\startdata
\multirow{9}{*}{\hfil 2005} & {\bf 7Px1sJu9\_07} & 11.387 & 137.259 & 8.678 & 11330.0 & 0.03119 & 2707.0 & 60.37 & good fit \\
 & ORya6h0F\_07 & 13.715 & 138.992 & 18.77 & 12710.0 & 0.01624 & 1362.0 & 68.83 & $L_{2005} \approx L_{2017}$  \\
 & e6RRFGdF\_04 & 15.246 & 143.522 & 8.979 & 17560.0 & 0.03541 & 1359.0 & 38.46 & $L_{2005} \approx L_{2017}$ \\
 & 1EHMCdxi\_07 & 15.652 & 135.411 & 17.09 & 13690.0 & 0.07064 & 4221.0 & 63.84 & $L_{2005} \approx L_{2017}$ \\
 & e6RRFGdF\_05 & 16.029 & 143.569 & 8.979 & 17560.0 & 0.03541 & 1359.0 & 41.11 & $L_{2005} \approx L_{2017}$ \\
 & {\bf wRz4n9OE\_04} & 16.72 & 120.122 & 7.19 & 10480.0 & 0.02166 & 3444.0 & 37.77 & good fit\\
 & {\bf LwIekVA6\_04} & 19.687 & 147.335 & 16.18 & 10510.0 & 0.03299 & 2554.0 & 31.59 & good fit \\
 & kskLMN4Y\_05 & 20.179 & 149.157 & 5.494 & 24540.0 & 0.0622 & 3640.0 & 47.86 & too hot \\
 & kskLMN4Y\_06 & 20.859 & 148.937 & 5.494 & 24540.0 & 0.0622 & 3640.0 & 51.65 & too hot\\
\hline
\multirow{13}{*}{\hfil 2017}  & {\bf 8UXTjAhl\_08} & 2.412 & 134.181 & 34.14 & 9631.0 & 0.07883 & 4067.0 & 72.06 & good fit\\
 & {\bf 8UXTjAhl\_07} & 7.531 & 134.866 & 34.14 & 9631.0 & 0.07883 & 4067.0 & 63.34 & good fit \\
 & yTvNrh2o\_04 & 10.444 & 117.76 & 4.789 & 22810.0 & 0.07566 & 316.5 & 32.15 & bad fit at $19.7,37.1\micron$\\
 & REwCcjZc\_05 & 10.863 & 132.046 & 39.91 & 8132.0 & 0.017 & 4539.0 & 48.53 & bad fit at $19.7,37.1\micron$ \\
 & REwCcjZc\_04 & 11.187 & 134.554 & 39.91 & 8132.0 & 0.017 & 4539.0 & 32.8 & bad fit at $19.7,37.1\micron$\\
 & yTvNrh2o\_05 & 11.204 & 119.925 & 4.789 & 22810.0 & 0.07566 & 316.5 & 42.77 & bad fit at $19.7,37.1\micron$ \\
 & REwCcjZc\_06 & 11.619 & 130.341 & 39.91 & 8132.0 & 0.017 & 4539.0 & 56.39 & bad fit at $19.7,37.1\micron$\\
 & REwCcjZc\_07 & 12.488 & 129.402 & 39.91 & 8132.0 & 0.017 & 4539.0 & 60.18 & bad fit at $19.7,37.1\micron$\\
 & V5ttV0Uz\_07 & 12.517 & 145.476 & 59.43 & 7264.0 & 0.0982 & 4549.0 & 67.08 & bad fit at $19.7,37.1\micron$ \\
 & h2UvYxug\_09 & 13.188 & 139.651 & 76.3 & 9036.0 & 0.01695 & 1770.0 & 82.7 & bad fit at $19.7,37.1\micron$ \\
 & LZEBw0DZ\_08 & 13.556 & 138.967 & 38.38 & 11690.0 & 0.01377 & 941.7 & 71.79 & bad fit at $19.7,37.1\micron$ \\
 & LZEBw0DZ\_07 & 14.313 & 139.677 & 38.38 & 11690.0 & 0.01377 & 941.7 & 60.78 & bad fit at $19.7,37.1\micron$\\
 & jzCvPxfT\_06 & 14.323 & 126.138 & 47.92 & 8185.0 & 0.0576 & 4076.0 & 52.93 & bad fit at $19.7,37.1\micron$ \\
\enddata
\end{deluxetable*}

The difference in luminosity between the 2005 and 2017 epochs indicates the luminosity burst of $\Delta L = 7600\pm900 L_\sun$ during an accretion outburst of M17~MIR. If $\Delta L$ is entirely attributed to increased accretion, we can predict the flux density of M17~MIR in the Spitzer I1 and I2 bands during the outburst. We assume that the blackbody temperature of $\Delta L$ during outburst is $1000\,\mathrm{K}$. Reddenning the blackbody radiation from $\Delta L$ with a certain amount of extinction essentially matches the flux density in the Spitzer I1 and I2 bands. We assume a total extinction to M17~MIR of 150\,mag, which is the sum of a foreground extinction of 130\,mag according to the good SED models for M17~MIR and an additional extinction of 20\,mag from the circumstellar disk. The blackbody radiation from $\Delta L=7600\pm900\,L_\sun$ suffers from a total extinction of 150\,mag, equivalent to $0.066\times150$\,mag and $0.052\times150$\,mag in the Spitzer I1 and I2 bands. This toy model yields flux density of 5.8 mJy and 52.5 mJy in the Spitzer I1 and I2 bands, respectively, which are comparable to the observed values $3.8\pm0.7$ mJy and $59.7\pm5.9$ mJy in the 2017 epoch, differing by a factor of less than 2.

The luminosities of M17~MIR in Table~\ref{tbl:lumin} are indeed the total luminosites, i.e., the combination of photospheric luminosity ($L_\ast$) and accretion luminosity ($L_\mathrm{acc}$),
  $$L = 4\pi R^2_\ast\sigma T^4_\mathrm{eff} + GM_\ast\dot M_\mathrm{acc}/R_\ast .$$
 In the 2005 epoch, we assume that half of $L_\mathrm{2005}=1400\,L_\sun$ arises from a ZAMS photosphere with a solar metalicity, and the progenitor's properties are: mass $M_\ast=5.4\,M_\sun$, radius $R_\ast=2.7\,R_\sun$, and effective temperature $T_\mathrm{eff}=18000\,\mathrm{K}$ \citep{1996MNRAS.281..257T}, which lies between spectral type B2.5V and B3V \citep{2013ApJS..208....9P}. The accretion rate $\dot M_\mathrm{acc}$ in the 2005 epoch needed to produce the half of $L_\mathrm{2005}$ is $\sim1.1\times10^{-5}\,M_\sun\,\mathrm{yr^{-1}}$, comparable to the steady accretion rates for the bursting MYSO of \citet{2021ApJ...912L..17H} and for the $6\,M_\sun$ formation model of \citet{2013A&A...557A.112H}. In the 2017 epoch, we assume $L_\mathrm{acc}\approx8300\,L_\sun$, which yields $\dot M_\mathrm{acc}\approx1.7\times10^{-3}\,M_\sun\,\mathrm{yr^{-1}} $ in the 2017 epoch, where $M_\ast=5.4\,M_\sun$ and $R_\ast=34.1\,R_\sun$ are assumed. The value of $\dot M_\mathrm{acc}$ in the 2017 epoch estimated from the stellar parameters agrees with the value of $\sim 2\times10^{-3}\,M_\sun\,\mathrm{yr^{-1}} $ obtained from the proper motions of H$_2$O maser spots (Sect.~\ref{Sect:rates}). In the 2017 epoch, a larger stellar radius is required to produce an accretion rate comparable to the value estimated from observations. This fact is consistent with the simulation prediction that a large accretion rate may cause the protostellar photosphere to bloat, resulting in larger radius \citep{2008ASPC..387..189Y, 2009ApJ...691..823H,2019MNRAS.484.2482M}.

\setcounter{table}{9}
\begin{table*}
\centering
\caption{Basic stellar parameters of M17~MIR in the 2005 and 2017 epochs.}
\label{tbl:lumin}
\begin{tabular}{c c c c c c c c }
  \hline\hline
  Epoch & $R_\ast$ & $ T_\mathrm{eff}$  & $ L_\ast$ & $A_V$  & $M_\ast$   & $\dot M_\mathrm{acc}$  & $L_{acc}$ \\
        & ($R_\sun$) & (K)   & ($L_\sun$) &  (mag) & ($M_\sun$)       &    ($M_\sun\,\mathrm{yr}^{-1}$) & ($L_\sun$) \\
  \hline
  2005 & 10.1(3.6)     & 10863(415) & 1400(897) & 134.7(10.4) & 5.4 & $1.1\times10^{-5}$ & $7.0\times10^2$ \\
  2017 & 34.1    & 9631 &  9034 & 134 & 5.4 & $1.7\times10^{-3}$ & $8.3\times10^3 $ \\
  \hline
\end{tabular}
\end{table*}


\setcounter{table}{10}
\begin{table*}
  \tabletypesize{\tiny}
\centering
\caption{Key parameters of six MYSOs with accretion Burst}
\label{tbl:com}
\begin{tabular}{c c c c c c c c c c}
  \hline\hline
  Name & $d$ & $L^\mathrm{pre}$  & $L^\mathrm{burst}$ & $L^\mathrm{post}$ & $\Delta F_\mathrm{IR}$ & $\Delta F_\mathrm{mm}$ &  $\Delta t$ &  Maser Flaring & Note \\
        & (kpc) & ($10^3L_\sun$)  & ($10^3L_\sun$) & ($10^3L_\sun$) &       &   &     (yr) &  & \\
  \hline
  V723 Car & 2.5     &  & - & 4 & $\sim10$ & -- & 5 & -- & $K\gtrsim12.9$ \\
  S255IR--NIRS3 & 1.8    & 29 & 160 & -- & $\gtrsim10$ & 2  & 2  & CH$_3$OH, H$_2$O & $K\gtrsim9$ \\
  NGC 6334I-MM1 & 1.3    & 2.9   & 47.6 & -- & 16.3 & $3.9\pm0.6$ & $>6$  & CH$_3$OH, H$_2$O & $K$-band invisible \\
  G323.46--0.08 & 4.8    & --   & -- & -- & -- & -- & --  & CH$_3$OH  & $K$-band visible\\
  G358.93--MM1 & 6.75 & 7.6 & 19.3 & 12.7 & $\gtrsim2$ & -- & $\lesssim0.5$ &  CH$_3$OH  & $K\gtrsim15$ \\
  M17~MIR      & 2.0  & 1.4  & 9.0  & 1.4 &  $\gtrsim10$ & --& $\sim2\times15$ &  H$_2$O & $K\gtrsim22$  \\
  \hline

\end{tabular}
\end{table*}

\subsection{Accretion History of M17~MIR}

The amplitude of MIR flux variation is directly proportional to the enhanced $L_\mathrm{acc}$. Nevertheless, the enhanced $L_\mathrm{acc}$ is a reduced version of enhanced $\dot M_\mathrm{acc}$, due to the significant inflation of the central object during the burst accretion phase. In the burst accretion phase, the $\dot M_\mathrm{acc}$ is amplified by a factor of hundreds compared with that in the quiescent phase. M17~MIR even shows a WISE/NEOWISE W2 flux variation in a cadence of six months, despite the overall trend of becoming continuously brighter. The overall trend is suggested to originate from the increasing $\dot M_\mathrm{acc}$. On the other hand, the smaller variation in the cadence of six months might imply unsteady $\dot M_\mathrm{acc}$ with a smaller amplitude. For instance, the variation in the W2 flux density is $-42\%$ between 2017.09 and 2018.03. This variation corresponds to $\Delta \dot M_\mathrm{acc}=-42\%$, if we assume the stellar radius to remain unchanged. This is the largest variation in $\dot M_\mathrm{acc}$ during the burst accretion phase.

Considering the MIR variation during the decreasing phase 1993.03--2004.09, M17~MIR is about three times brighter in 1993.03 than in 2017.08. We speculate that $\dot M_\mathrm{acc}$ in 1993.03 is $\sim5\times10^{-3}\,M_\sun\,\mathrm{yr}^{-1}$, three times that in the rebrightening phase. However, $\dot M_\mathrm{acc}$ in 1993.03 is probably even higher, since M17~MIR may expand to even larger size than that in the rebrightening phase.

The MIR light curve of M17~MIR exhibits three phases characterized by M17~MIR's MIR variation in flux density over 26 yr. From the above results and analyses, we conclude that the decreasing and rebrightening phase of M17~MIR indeed reflect the two accretion bursts, which are separated by the quiescent phase lasting about 6 yr. The first accretion burst likely began between 1984 and 1993, and lasted longer than 11 yr but less than 20 yr. Thus we use a middle value of 15 yr. We fortunately caught the initial stage of the second accretion burst, aided by the multiepoch observations of the Spitzer and WISE/NEOWISE space telescopes. The duration of the second accretion burst is definitely longer than 9 yr. It is very likely that M17~MIR has not yet reached to its peak brightness, and the duration of the second accretion burst is similar to that of the first epoch. M17~MIR is the first MYSO showing multiple accretion bursts with durations of tens of years and a burst magnitude of about 2\,mag. The observed properties of the accretion bursts of M17~MIR agree with the prediction from theoretical simulation that 2\,mag bursts have a mean duration of about 20\,yr \citep{2019MNRAS.482.5459M}.

We assume a bursting accretion timescale of 15 yr, which is comparable to the duration of the decreasing phase in the MIR light curve (somewhere between 1984 and 1993 to mid-2004). If M17~MIR can sustain this enhanced disk accretion rate of $\sim2\times10^{-3}\,M_\odot\,\mathrm{yr^{-1}}$, it could gain about $0.03\,M_\odot$ during one accretion burst. This amount of accreted mass is an order of magnitude higher than the mass accretion onto S255IR--NIRS3 \citep{2017NatPh..13..276C} and two orders higher than that onto G358.93--MM1 \citep{2021A&A...646A.161S}. The quiescent phase of M17~MIR lasts about 6\,yr. During the time span of 26 yr, the fraction of time in accretion burst is about 83\%. The main accretion phase of a high-mass protostellar object in an HMC lasts a few times $10^4$ yr \citep{2005ApJ...624..827F,2017A&A...604A..60B}. A crude estimate, assuming $t_\mathrm{acc}=1\times10^{4}\,\mathrm{yr}$, is that this fraction of time in accretion burst yields $\sim14\,M_\odot$ mass that M17~MIR could accrete from its circumstellar disk. 

\subsection{Comparisons to other variable MYSOs }
The sample of MYSO accretion bursts is still small. Six MYSOs show a luminosity increase due to an accretion burst: V723 Car \citep{2015MNRAS.446.4088T}, S255IR--NIRS3 \citep{2017NatPh..13..276C,2018ApJ...863L..12L}, NGC 6334I--MM1 \citep{2017ApJ...837L..29H,2021ApJ...912L..17H}, G323.46--0.08 \citep{2019MNRAS.487.2407P}, G358.93--0.03 MM1 \citep{2021A&A...646A.161S}, and M17~MIR in the present work. However, these MYSOs are in different pre-main-sequence (PMS) evolutionary stages. The durations of accretion bursts are poorly constrained except for M17~MIR. To qualitatively describe the relation between the PMS evolutionary stage and an accretion burst, their key parameters are summarized in Table~\ref{tbl:com}. Relatively evolved MYSOs are commonly brighter in the near-IR $K$ band, while the earliest MYSOs are very faint or even invisible in the same band. Generally the $K$-band brightness of these MYSOs is representative of the PMS evolutionary stage. Among the six MYSOs tabulated in Table~\ref{tbl:com}, four sources show variabilities in the $K$ band, indicating relatively late PMS stages. In particular, G323.46--0.08 was the first one that showed periodic 6.7\,GHz methanol maser emission with a period of about 90 days \citep{2019MNRAS.487.2407P}. Later -- on the basis of multiepoch IR data -- it was confirmed as an accretion burst MYSO \citep[see][and references therein]{2021A&A...646A.161S}.

For V723 Car, S255IR--NIRS3, and G358.93--MM1 the duration of the accretion burst $\Delta t$ was constrained from their IR luminosity changes. The durations of their luminosity increases are of order of years and thus comparable to each other. The two earliest MYSOs, NGC 6334I--MM1 and M17~MIR, are both embedded in HMCs. The progenitor of NGC 6334I-MM1 in its pre-burst phase might be able to produce a hypercompact region MM1B \citep{2016ApJ...832..187B}, implying that NGC 6334I-MM1 is more massive and at a later evolutionary stage than M17~MIR. Acoording to the schematic evolutionary sequence of a massive protostar \citep{2018ARA&A..56...41M}, M17~MIR might be the earliest MYSO with accretion outbursts. Considering the extreme youth of M17~MIR, its multiple accretion bursts suggest that minor accretion bursts are frequent in the very early stages of massive star formation.

\section{Conclusions}\label{Sect:Con}

The multiwavelength and multiepoch data of the massive protostar M17~MIR and the subsequent analyses of these data yield the following major results.

1. By analyzing the IR data of M17~SW in the $3-70\,\mu$m range, M17~MIR is identified as the driving source of the expanding structure traced by H$_2$O maser motions. The sub-mm dust continuum emission and molecular line emission indicates that M17~MIR is embedded within a HMC.

2. The unified flux density of M17~MIR at the $3.8\,\mu$m $L'$ band and the $4.6\,\mu$m WISE W2 band clearly shows three stages in the light curve: i) the decreasing phase during 1993.03 to mid 2004 with an amplitude of 10 at $3.8\,\mu$m, ii) the quiescent phase during mid 2004 to mid 2010, and iii) the rebrightening phase since mid 2010 until now with an amplitude of 5 at $4.6\,\mu$m. The 22\,GHz H$_2$O maser emission associated with M17~MIR also shows flux variations. They are correlated with the MIR flux variation in time space, which together indicate enhanced disk accretion rate onto M17~MIR during the decreasing and rebrightening phase.

3. The kinematics of H$_2$O maser spots, measured in 2009.10--2010.12, in the close vicinity of M17~MIR yield a disk accretion rate $\dot M_\mathrm{acc}\sim7\times10^{-4}\,M_\sun\,\mathrm{yr^{-1}}$. In the later stage of the rebrightening phase after 2014, $\dot M_\mathrm{acc}$ is $\sim2\times10^{-3}\,M_\sun\,\mathrm{yr^{-1}}$, given by the $4.6\,\mu$m flux increased by a factor of $\sim 3$ compared to the initial stage in 2010.

4. Radiative transfer modeling for the SEDs in the 2005 and 2017 epoch constrains the basic stellar parameters of M17~MIR. In the quiescent phase in the 2005 epoch, M17~MIR likely has total luminosity $L_{2005}=1400\pm897\,L_\sun$, stellar mass $M_\ast\sim5.4\,M_\sun$, stellar radius $R_\ast=10.1\pm3.6\,R_\sun$, and accretion rate $\dot M_\mathrm{acc}\sim1.1\times10^{-5}\,M_\sun\,\mathrm{yr}^{-1}$. In the accretion burst phase (e.g. in the 2017 epoch),  $L_{2017}=9034\,L_\sun$( $\Delta L\approx7600\,L_\sun$), $R_\ast=34.1\,R_\sun$, and $\dot M_\mathrm{acc}\sim1.7\times10^{-3}\,M_\sun\,\mathrm{yr}^{-1}$.  In the accretion outburst, a larger stellar radius is required to produce $\dot M_\mathrm{acc}$ consistent with the value estimated from the kinematics of H$_2$O maser spots.

5. The decreasing and rebrightening phase of M17~MIR reflect two accretion bursts ($\Delta t\sim 9-20$ yr), which are separated by the quiescent phase lasting about 6\,yr. During the time span of 26 yr, the fraction time of accretion burst is about 83\%. M17~MIR is the first discovered massive protostar showing multiple accretion bursts with durations of tens of years and a burst magnitude of about 2\,mag. Among the limited sample of accretion burst MYSOs, the extreme youth of M17~MIR suggest that minor accretion bursts are frequent at very early stages of massive star formation.

6. Even during the current accretion burst (since mid 2010 to until now), the W2 flux density of M17~MIR shows variations of small amplitude up to a factor of 3, indicating variable accretion rate with amplitutde much lower than the overall contrast between queiscent and outburst state. Long time monitoring at MIR bands will record the bursting accretion activity of M17~MIR in the near future. Because M17~MIR is still in its earliest stage of evolution, it will be a unique testbed for probing the earliest phases of massive star formation.

\begin{acknowledgements}

The authors would like to thank the anonymous referee for a detailed and thoughtful review that has improved the scientific contents of the paper. This work is supported by National Key Research \& Development Program of China (2017YFA0402702). We acknowledge support from the general grants 11903083, U2031202, 11873093, 11873094, and 12041305 of National Natural Science Foundation of China, and support from the DFG grant CH 71/33-1. We thank Tie Liu for informing us of the JCMT SCUBA2 observations in 2016 April that are used in this research. We thank J. O. Chibueze for the discussion of VERA observations. This research has made use of the observations performed with the Herschel Space Observatory, which is an ESA space observatory with science instruments provided by European-led Principal Investigator consortia and with important participation from NASA. This research has made use of the NASA/IPAC Infrared Science Archive, which is operated by the Jet Propulsion Laboratory, California Institute of Technology, under contract with the National Aeronautics and Space Administration. This work is based in part on observations made with the Spitzer Space Telescope, which is operated by the Jet Propulsion Laboratory, California Institute of Technology under a contract with NASA. This publication makes use of data products from the Near-Earth Object Wide-field Infrared Survey Explorer (NEOWISE), which is a joint project of the Jet Propulsion Laboratory/California Institute of Technology and the University of Arizona. NEOWISE is funded by the National Aeronautics and Space Administration. This research is based in part on observations made with the NASA/DLR Stratospheric Observatory for Infrared Astronomy (SOFIA). SOFIA is jointly operated by the Universities Space Research Association, Inc. (USRA), under NASA contract NNA17BF53C, and the Deutsches SOFIA Institut (DSI) under DLR contract 50 OK 0901 to the University of Stuttgart. This research used the facilities of the Canadian Astronomy Data Centre operated by the National Research Council of Canada with the support of the Canadian Space Agency. This research has made use of the SVO Filter Profile Service (http://svo2.cab.inta-csic.es/theory/fps/) supported from the Spanish MINECO through grant AyA2014-55216. This research has made use of NASA's Astrophysics Data System.
\end{acknowledgements}

\bibliographystyle{aasjournal}
\bibliography{myrefs}

\end{CJK*}
\end{document}